\documentclass[9pt,twocolumn,twoside]{pnas-new}
\pdfoutput=1
\usepackage{url}
\usepackage{pdfpages}
\usepackage{import}

\templatetype{pnasresearcharticle} 

\title{Temporal Limits of Privacy in Human Behavior}

\author[a,b,1,*]{Vedran Sekara}
\author[a, b]{Enys Mones} 
\author[a,c,1,*]{H\r{a}kan Jonsson}

\affil[a]{Sony Mobile Communications, SE-22188 Lund, Sweden}
\affil[b]{Department of Applied Mathematics and Computer Science, Technical University of Denmark, DK-2800 Kongens Lyngby, Denmark}
\affil[c]{Faculty of Engineering (LTH), University of Lund, SE-22100 Lund, Sweden}

\leadauthor{Sekara} 

\authorcontributions{V.S. and H.J. conceived the study.  V.S. and H.J designed measures and analyses. V.S. collected and curated the data. V.S. conducted the analysis. V.S., E.M., and H.J. wrote the manuscript. All authors interpreted and discussed the findings.}
\authordeclaration{The authors have no conflict of interest.}
\equalauthors{\textsuperscript{1}V.S. and H.J. lead the research and contributed equally to this work.}
\correspondingauthor{\textsuperscript{*}To whom correspondence should be addressed: hajons@gmail.com, vedransekara@gmail.com.}


\begin{abstract}
Large-scale collection of human behavioral data by companies raises serious privacy concerns. We show that behavior captured in the form of application usage data collected from smartphones is highly unique even in very large datasets encompassing millions of individuals. This makes behavior-based re-identification of users across datasets possible. We study 12 months of data from 3.5 million users and show that four apps are enough to uniquely re-identify 91.2\% of users using a simple strategy based on public information. Furthermore, we show that there is seasonal variability in uniqueness and that application usage fingerprints drift over time at an average constant rate.
\end{abstract}

\begin{document}

\verticaladjustment{-2pt}

\maketitle
\thispagestyle{firststyle}
\ifthenelse{\boolean{shortarticle}}{\ifthenelse{\boolean{singlecolumn}}{\abscontentformatted}{\abscontent}}{}

\dropcap{T}racking behavior is a fundamental part of the emerging big-data economy, allowing companies and organizations to segment, profile and understand their users in increasingly greater detail.
Modeling context and interests of users has proven to have various advantages: products can be designed to better fit customers' needs; content can be adapted; and advertising can be made more relevant~\cite{agrawal1993mining,bell2007lessons,chen2009large,mislove2010you,dodds2010measuring,mislove2011understanding}.
Efficient user modeling requires the collection of large-scale datasets of human behavior, which has led to a growing proportion of human activities to be recorded and stored~\cite{conte2012manifesto}.
Today, most of our interactions with computers are stored in a database, whether it is an e-mail, phone call, credit-card transaction, Facebook like, or online search, and the rate of information growth is expected to accelerate even further in the future~\cite{lazer2009computational}.
These rich digital traces can be compiled into detailed representations of human behavior and can revolutionize how we organize our societies, fight diseases, and perform research; however, they also raise serious privacy concerns~\cite{blumberg2009locational,eckersley2010unique,de2013unique,hannak2013measuring,greenwood2014new,de2015unique,sapiezynski2015tracking,mayer2016evaluating}.
For example, Narayanan et al. demonstrated the feasibility of inferring political views of IMDb users through re-identification of  movie ratings~\cite{narayanan2008robust}.
Another infamous case is the hacking (and eventual erasure of personal data) of multiple accounts of a journalist, which was carried out by the attacker being able to connect two different databases~\cite{wiredhack}.

The ubiquity and sensing capabilities of mobile phones together with our seemingly symbiotic relationship to them, renders these devices good tools for tracking and studying human behavior~\cite{eagle2006reality,stopczynski2014measuring}.
Mobile phones are ubiquitous and have permeated nearly every human society: in the year 2015 98.3\% of the world's population had a mobile subscription~\cite{itu2016}.
Mobile phones have transformed the way people access the internet as well: today the majority of traffic to web pages stems from mobile devices rather than from desktop computers~\cite{stonetemple}, making advertisers target mobile phones to a higher degree.
With the standard methods based on cookies for identifying customers not being used in smartphone apps, along with the rising usage of ad-blockers among users~\cite{pagefair2017}, advertisers and so-called \emph{data brokers} are now targeting smartphone applications to replace the rich data cookies provided in the past.
Advertisement identifiers are one such ID embedded in applications, but they do not allow data brokers to track users across multiple applications or devices, and they can even be reset by the user.
Application usage behavior, however, cannot be cleared, and it is hard (and in many cases not feasible) to be changed or manipulated by users.
This creates an economic incentive for global population tracking of application usage.
This tracking is in conflict of users' perception of permissible usage of data~\cite{martin2018penalty}.
Also, in general, users are not knowledgeable enough about what data is collected about them to make an informed decision~\cite{posner1981economics}.

A majority of the online services people interact with on a daily basis collect personal information and sell the data to data brokers (third parties)~\cite{anthes2015data}.
In a recent report released by the U.S. Federal Trade Commission, it was shown that data broker companies obtain vast amounts of personal data, which they further enrich with additional online and offline sources, and re-sell these improved datasets to the highest bidder, typically without the explicit consent or knowledge of the users~\cite{ramirez2014data}.
According to U.S. privacy laws, data is considered anonymous if it does not contain personally identifiable information (PII) such as name, home address, email address, phone number, social security number, or any other obvious identifier.
As a result, it is legal for companies to share and sell anonymized versions of a dataset.
However, as studies have shown, the mere absence of PII in a dataset does not necessarily guarantee anonymity due to the fact that it is relatively easy to compromise the privacy of individuals~\cite{narayanan2008robust,sweeney2002k,de2013unique}.

Human behavior, although imbued with routines, is inherently diverse.
Previous work has shown that 99.4\% of smartphone users have unique app usage patterns and established the viability of using apps as markers of human identity, similar in application to fingerprints in forensic science~\cite{falaki2010diversity,welke2016differentiating,achara2015unicity}.
It has further been demonstrated that the software infrastructure we use to access the Internet can be used to identify users~\cite{eckersley2010unique}.
The digital breadcrumbs we leave online can be used to infer many aspects of our lives.
It has been shown for example that age, gender, relationship status, education level, political beliefs, sexual orientation, religion, and even personality can be predicted from Facebook likes~\cite{kosinski2013private,youyou2015computer}, or based on the apps people use on their smartphones~\cite{chittaranjan2013mining,seneviratne2014predicting,malmi2016you}.
Human mobility traces has been shown to be highly unique and research has further shown that 4 spatio-temporal points are sufficient to re-identify a majority of individuals~\cite{de2013unique}.

This study demonstrates how easy it is to uniquely identify individuals from their smartphone usage patterns given only a handful of data points, and investigates the temporal patterns of uniqueness, revealing that humans are easier to identify during certain periods of the year.
We define identification as matching a behavior pattern against an (anonymous) quasi-identifier consisting of a similar pattern.
In the dataset we use, no further information can be gained about the user beyond matching two patterns.
However, in a real world scenario, an attacker could use this method for connecting two datasets to learn new information about the re-identified user, e.g. email address, age, or gender, depending on the data available to the attacker.
Our study focuses on applications (apps) --- small software programs which users can download to their smartphones, and which provide a near unlimited range of functions, from simple functions such as flashlights or calculators to more advanced---artificial intelligence like---functions.
Each new phone comes with a set of apps pre-loaded by the manufacturer, but a user is free to customize their device to suit their specific needs, as such users have access to millions of apps on app stores such as \emph{Google Play} (approx. 2.8 million apps)~\cite{appbrain}.

\section*{Results}
\subsection*{Uniqueness of human behavior}
To evaluate the likelihood of identifying individuals within smartphone usage data we use a dataset that spans 12 months (Feb. 1st 2016 to Jan. 31st 2017) and encompasses 3.5 million people using in total 1.1 million unique apps.
We have chosen to disregard phone vendor specific apps, such as alarm clock apps, built-in phone dialer apps, etc. and only focus on apps that are downloadable from Google Play. From this we form app \emph{fingerprints} for each user, i.e. a binary vector containing information about which apps the user has used for every month. We only consider apps actually used by a user in a month, not apps that were installed but never used.
Figure 1 illustrates the typical patterns of app usage, with individuals continuously changing their app-fingerprint over the course of a year by trying out new apps and ceasing to use others.
As such, app-fingerprints slowly drift over time, with the average rate of change being roughly constant between consecutive months (Figure S1).
In combination with fingerprints drifting, the number of apps people use on their smartphones is constant over time as well, suggesting that humans have a limited capacity for interacting, navigating, and managing the plethora of services and social networks offered by smartphones (Figure S2).
This limiting effect has been observed in other aspects of life such as interactions among people~\cite{dunbar1992neocortex} or geo-spatial exploration~\cite{alessandretti2016evidence}.

\begin{figure}[!hptb]
  \centering
    \includegraphics[width=0.49\textwidth]{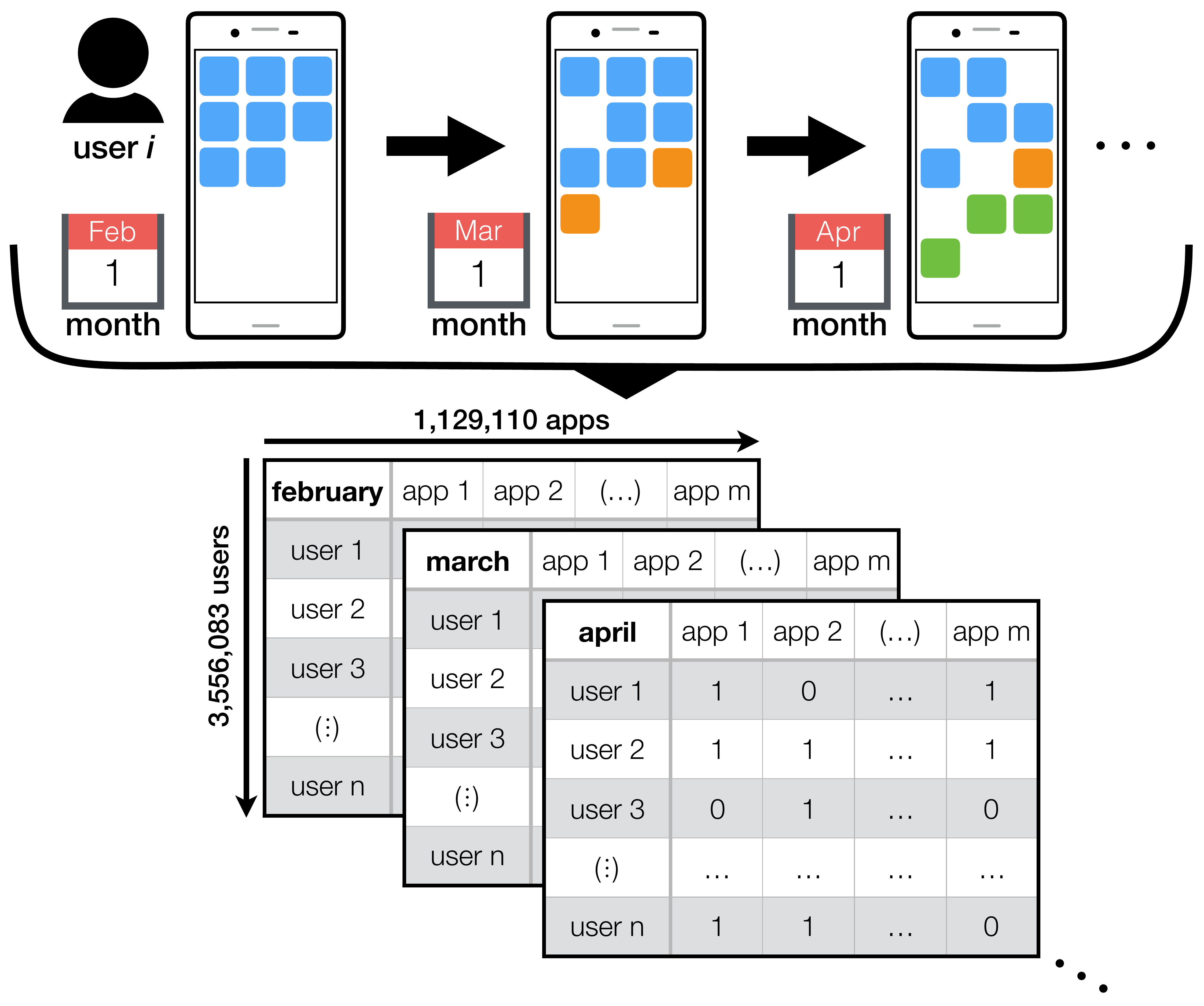}
    \label{fig:illustration}
    \caption{Smartphone usage patterns change over time, with users continuously changing which apps they use.
    This study is based on smartphone app-fingerprints of 3,556,083 individuals.
    For each month between February 2016 and January 2017, we retrieve the list of apps a person has used during the period ($n_{\text{month}} = 23$ apps per person per month on average, or $n_{\text{year}} = 76$ apps on average during the full 12-month period).
    App-fingerprints are represented as a sparse \textit{user} $\times$ \textit{app} $\times$ \textit{month} tensor, with $1$ indicating a person has used an app during a specific month, $0$ otherwise.
    To look at longer time-windows, we aggregate entries according to a maximum value heuristic and retain entries if they are greater than zero.
}
\end{figure}

The risk of re-identifying individuals is estimated by means of unicity~\cite{de2013unique,de2015unique}.
Here, re-identification corresponds to successful assignment of an app-fingerprint to a single unique user in our dataset.
This does not entail that we can directly get the \textit{real} identity of a person, such as name, address, e-mail, social security number, etc. 
This, however, would become possible if this knowledge is cross-referenced with other data sources, which there unfortunately has been countless examples of~\cite{narayanan2008robust,barbaro2006face,barth2012re,sweeney2013identifying,tockar2014riding}.
Given an individual's app-fingerprint, unicity quantifies the number of apps needed to uniquely re-identify that person; the fewer apps we need the more unique a person is and vice versa.
Given a dataset of app-fingerprints and set of apps $i$, $j$ and $k$, a user $u$ is uniquely identifiable if that user, and only that user, in the dataset has used apps $i$, $j$ and $k$, i.e. matching the fingerprint of user $u$.
In our dataset we evaluate uniqueness as the percentage of users we can re-identify using $n$ number of apps.

To attack the dataset without any prior knowledge of the system itself, the most realistic strategy is to pick apps at random.
Figure 2A shows the efficiency of this type of random sampling of apps, with $21.8\%$ of users being re-identified from using 4 apps.
Although this value means only 1 of every 5 individual can be re-identified, it is surprisingly high given that we only use binary features (that is, has the user used the app or not) and have no information regarding \emph{when} an app was used or for \emph{how long}---features which would only make fingerprints more unique.
In case of a real attack, however, the above results might give the general public a false sense of security as it is possible to use free, publicly available information to formulate an attack strategy that greatly outperforms the random strategy.

The popularity of apps follows a heavy-tailed distribution~\cite{olmstead2016apps} (and see Figure S3); a few apps are used by millions or even billions of individuals, while an overwhelming majority of apps only have a couple of users.
All this information is available on \emph{Google Play} from where it is possible to retrieve by automatic means, or it can be purchased from vendors such as AppMonsta.
Because this information is so easily attainable, we formulate a strategy that takes the user base of apps (popularity of apps) into account, starting with the least used apps: the \textit{popularity strategy}.
Rather than using the popularity in terms of downloads on Google Play, we use the popularity counted as the number of users that use an app in our dataset (see Methods for details).
A real-world re-identification attack strategy could use the Google Play download numbers for each app to reduce the amount of computation required.
Figure 2B shows that just using 2 apps with the popularity strategy greatly outperforms the random strategy, and using 4 apps, we are able to re-identify $91.2\%$ of users.

\begin{figure}[!hptb]
  \centering
    \includegraphics[width=0.49\textwidth]{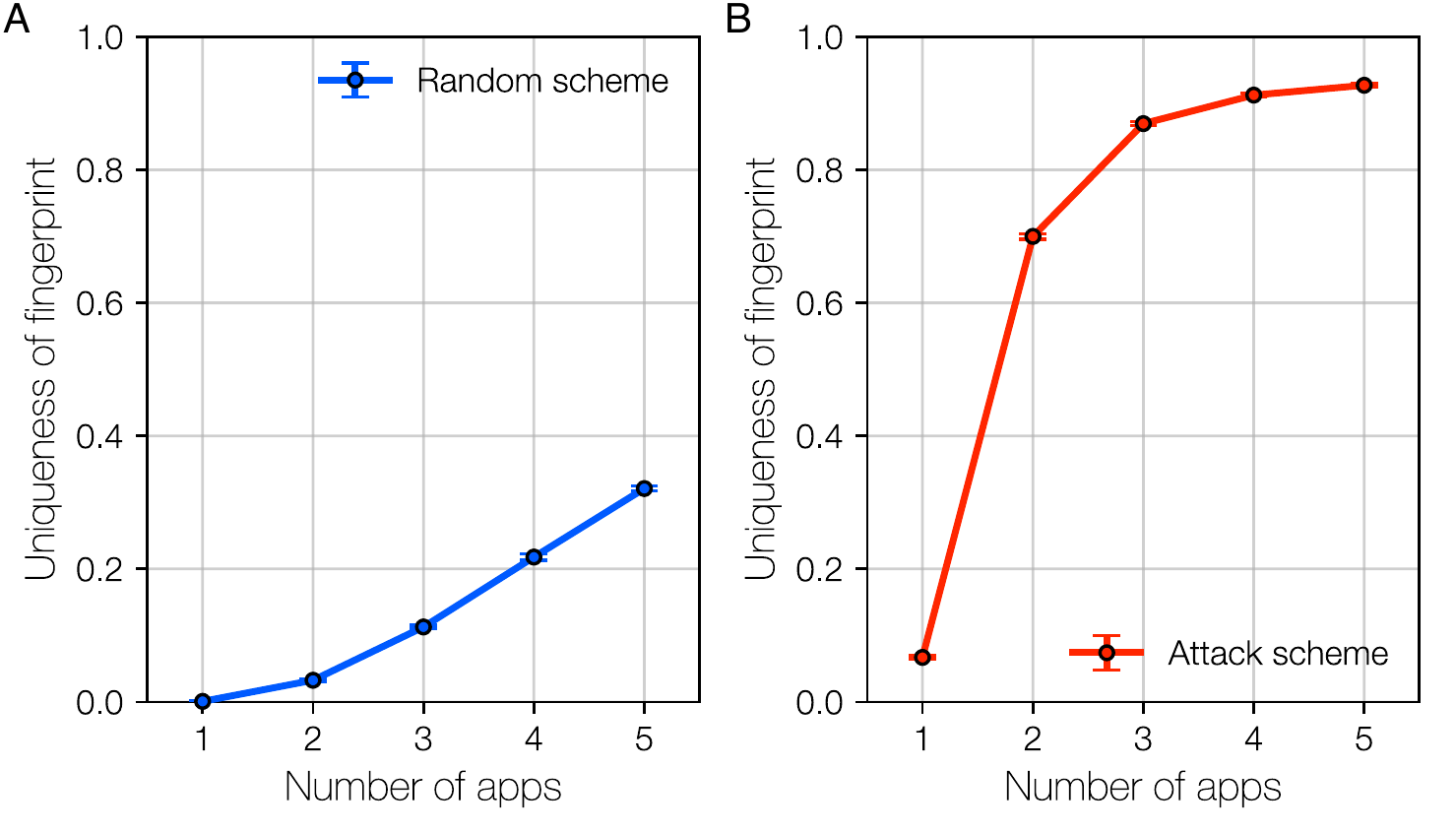}
    \label{fig:unicity}
    \caption{Uniqueness of smartphone app-fingerprints given $n$ number of apps. (A) Selecting apps at random is not an efficient way of identifying individuals and achieves a modest re-identification rate of $21.8\%$ when using 4 apps. (B) Using freely available outside information from Google Play to attack the problem yields significantly higher rates of re-identifications, $91.2\%$ when using 4 apps. Error bars denote one standard deviation.
    App-fingerprints are constructed from the full 12 months of data, and $99.7\%$ of individuals within our dataset have a unique fingerprint.}
\end{figure}

\subsection*{Seasonal variability of anonymity}
Human lives, routines and behaviors evolve over time~\cite{kossinets2006empirical,sekara2016fundamental,alessandretti2016evidence}, and therefore individual app-fingerprints might become harder (or easier) to identify.
To quantify the seasonal variability of uniqueness, we construct monthly fingerprints for all individuals and evaluate anonymity using the unicity framework.
Figure 3 shows the fraction of individuals that are re-identifiable per month, and reveals an increased fraction of identifications for June, July, and August---months which are typically considered vacation months.
The increase in uniqueness is independent of how we select apps (random, or by popularity).
In fact, during these three months the process of identifying individuals from randomly selected apps is respectively $14.8\%$ and $18.4\%$ more effective when using $5$ and $10$ apps.
For the popularity scheme, we note $6.8\%$ and $8.0\%$ higher rates of identifications when using $5$ and $10$ apps.
The increase in identifiability stems from a combination of related behavioral changes (Figure S4). 
Apps related to categories such as travel, weather, sports, and health \& fitness gain popularity during the summer months (June, July, August), related to people traveling and downloading apps that help them navigate new cities, using fitness apps to motivate them to exercise more, and using apps that enable them to follow global sports events such as the 2016 UEFA European Championship in football (soccer).
Simultaneously, apps related to categories such as education and business become less popular.  
This suggests an interplay between our physical behavior and our app-fingerprint, indicating that when we change our geo-spatial routines by traveling and exploring new places, we also change our app usage. This change in phone behavior makes our app-fingerprints more unique and easier to identify.

\begin{figure}[!hptb]
  \centering
    \includegraphics[width=0.49\textwidth]{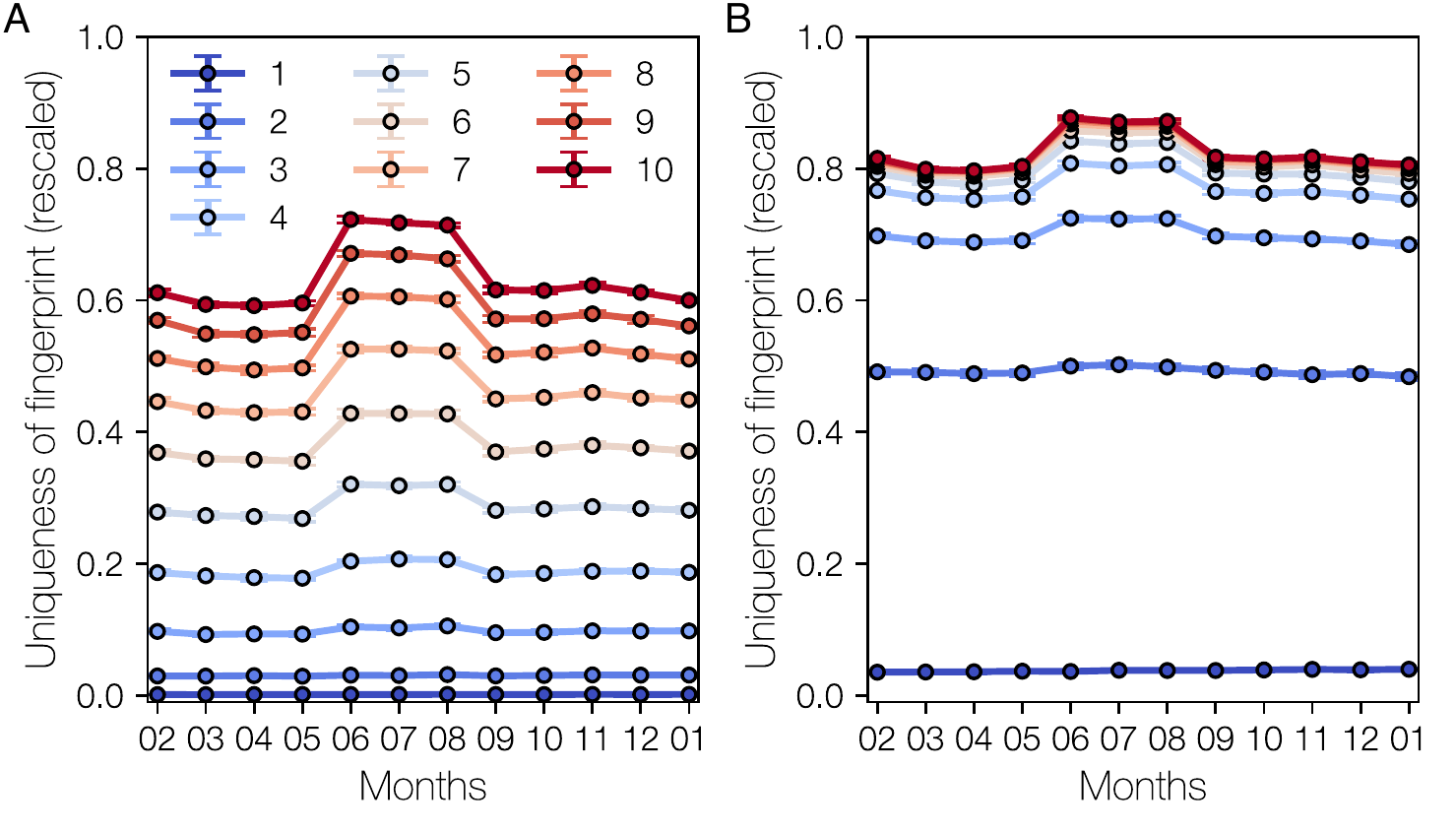}
    \label{fig:unicity_time}
    \caption{Seasonal variations of re-identifiable app-fingerprints over 12 months.
    The fraction of individuals which we can re-identify by using $n$ apps (1-10) changes from month to month, revealing that uniqueness has a temporal component, and that people are more unique during summer.
    This is independent of whether apps are selected using: (A) a random heuristic or (B) an attack scheme.
    Compared to Figure 2, the fraction of re-identified individuals per month is lower because we have segmented behavior into monthly fingerprints as compared to constructing fingerprints from 12 months of data.
    Uniqueness is rescaled according to the set size of apps present within each month (see Figure S5).
    }
\end{figure}

\subsection*{Hiding in the crowd}
Our dataset is limited to 3.5 million users, similar in size to a small country, but how will uniqueness change as more users are added (increased sample-size)?
Will it become possible to hide in the crowd?
More precisely, how does the population size affect the extent to which a specific app-fingerprint remains unique.
That is, as more and more users are added to our sample, does the likelihood to observe multiple individuals with identical fingerprints also increase?
This corresponds to an inverse k-anonymity problem~\cite{sweeney2002k}, where one needs to estimate the number of users that should be added in order to increase the overall anonymity of the dataset.
(Bearing in mind that overall anonymity is not a good measure for the sensitivity of individual traces.)
To understand the effect of sample-size on unicity, we first slice our dataset into smaller subsamples and use it to estimate the uniqueness for sample sizes ranging from 100,000 to 3.5 million individuals.
Figure 4A reveals that sample size has a large effect on the re-identification rate when selecting apps using a random heuristic.
Considering $n_{\text{apps}} = 5$, the average re-identification rate decreases from $45.89\%$ for a sample size of 1 million individuals to $37.33\%$ for 2 million individuals and $32.09\%$ for the full sample of 3.5 million people.
The attack scheme is considerably less affected (Figure 4B).
For $n_{\text{apps}} = 5$ we find that the re-identification rates are respectively $96.60\%$, $94.23\%$ and $92.72\%$ for sample sizes of 1, 2 and 3.5 million individuals. 
As such, increasing the sample size by $250\%$ (from 1 to 3.5 million individuals) only reduces uniqueness by approximately 4 percent-points.

In order to estimate uniqueness for sample sizes larger than the study population we extrapolate results from Figure~4B for $n_{\text{apps}} = 5$.
We express uniqueness of fingerprints using multiple functional forms including: power-laws ($\sim x^{\gamma} $), exponentials ($\sim \exp(\gamma x)$), stretched exponentials ($\sim \exp(x^\gamma)$), and linear functions ($\sim x$), where $x$ denotes the sample size and $\gamma$ is a scaling factor. 
The stretched exponential and power-law show the highest agreement with the data (Figure S6), and roughly suggest that 5 apps are enough to re-identify 75\%--80\% of individuals for 10 times larger samples (35 million individuals). 
Although the applied analysis displays high uncertainty with regards to extrapolations, it illustrates the observation that increasing the population size does not help us in hiding in the crowd (that is, uniqueness is not a characteristic of small sample sizes).

\begin{figure}[!hptb]
  \centering
    \includegraphics[width=0.49\textwidth]{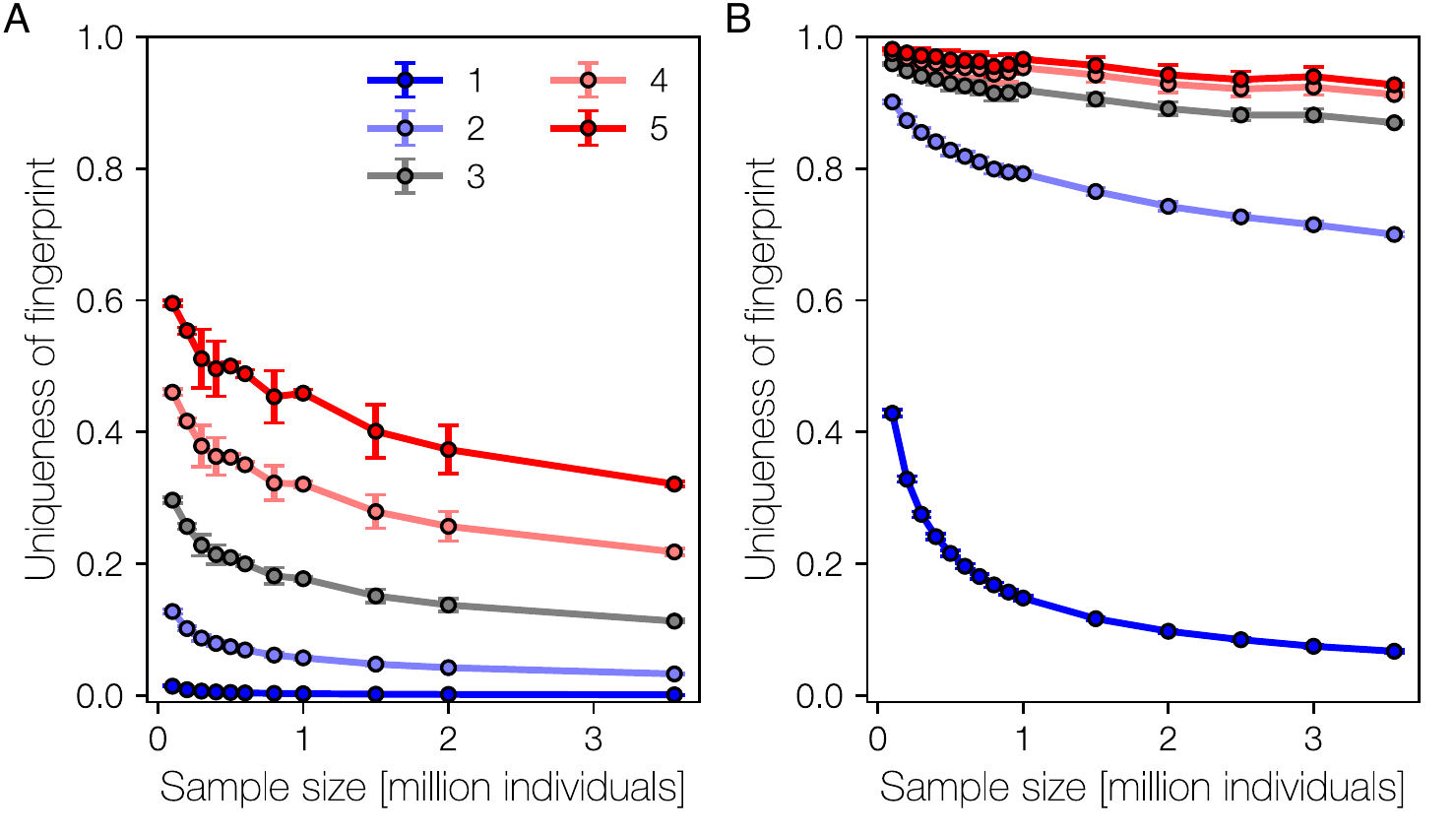}
    \label{fig:unicity_sample-size}
    \caption{Identifying fingerprints across data-samples with varying population sizes.
    Fingerprints are constructed from 12 months of data.
    The uniqueness of individual fingerprints is reduced (lower re-identification rates) as we increase the sample-size independently of whether apps are selected: (A) randomly or (B) according to the attack heuristic.
    The magnitude of the change, however, varies greatly different between the two heuristics.
    Results show in both panels are calculated from multiple realizations of the data (see Materials and Methods section).
    }
\end{figure}

\section*{Discussion}
Phone behavior is different from credit card traces and mobile phone mobility data in that the ease with which data can be collected, and any Android app can request permission to access your app history. We reviewed apps with more than 100,000 downloads which request the `retrieve running apps' permission on Android, and that are free (no price or in app purchases). Out of these 40 apps 31 contain ads. There are 15 apps that belong to the Personalization or Tools category, mostly anti-virus or launcher apps, which may need the permission to provide their features. For the other 25 apps, we found no features in the app that would motivate requesting this permission. Some of these apps are from major phone vendors whose privacy policy says they may share data with third parties.

The economic incentives, the easy and global scale of collecting and trading this data without users' knowledge creates some serious concerns, especially since this practice is in violation of users' expectations or knowledge~\cite{martin2018penalty, posner1981economics}.
The EU General Data Protection Regulation~(GDPR) may be a first step towards addressing these concerns through regulation, since it does mention unicity~\cite{gdpr} and applies globally to data about any EU citizen.
Our conclusion from this study is that application usage data should be considered personal information, since it is a unique fingerprint.

This study was performed using app usage data collected from Android phones from a single vendor only.
As phone vendor specific apps were disregarded in the analysis, we expect the results to generalize across all Android devices.
Further, we have no reason to believe that app usage behaviour and uniqueness is fundamentally different for individuals using iOS devices compared to Android users.

\matmethods{

\subsection*{The dataset}
We use a dataset that spans 12 months, from Feb. 1st 2016 to Feb. 1st 2017, and contains monthly app-fingerprints for 3,556,083 individuals with pseudonymized app and user identifiers.
Each fingerprint is a binary vector composed of the apps a person has used during a month. We do not consider apps that are installed but unused.
We further disregard phone vendor specific apps such as: alarm clock, phone dialer, settings etc. and only focus on apps that are downloadable from Google Play.
This removes vendor bias, and makes re-identification harder. The users are selected from major markets in the Americas, Europe and Asia. Thus, the impact of regional variations on uniqueness due to local applications is smaller than if we had sampled users from anywhere in the world.
In total, the number of unique apps in the dataset is 1,129,110, and each individual in the dataset uses at least 3 apps per month. 
Data collection is approved by the Sony Mobile Logging Board and written consent in electronic form has been obtained for all study participants according to the Sony Mobile Application Terms of Service and the Sony Mobile Privacy Policy.
Raw data cannot be shared publicly on the web, but we offer the possibility to reproduce our results starting from raw records by spending a research visit at Sony Mobile Communications.

\subsection*{Estimating uniqueness} 
To estimate the uniqueness of app-fingerprints, we apply the unicity framework~\cite{de2013unique} on $k$ samples of 10,000 randomly selected individuals.
For each individual we select $n$ apps (without replacement) from the person's app-fingerprint.
With the popularity based attack, apps with low user base are selected to increase the uniqueness of the app usage pattern.
The person is then said to be unique if they are the only individual in the dataset whose app-fingerprint contains those apps.
In cases where $n$ is larger the the total length of a person's app-fingerprint we instead select $\min(n,|\text{fingerprint}|)$ number of apps.
Uniqueness for a sample $k_i$ is then estimated as the fraction of the users that have unique traces.
Overall uniqueness is the average of the $s$ samples, and error-bars are given by the standard deviation.
We use $s=20$.

\subsection*{Subsampling the dataset}
To quantify the relation between sample size and uniqueness, we subsample the dataset by selecting a fraction of the original dataset.
For each sample $s_i$ we estimate uniqueness using the above methodology.
To account for selection bias we estimate uniqueness as the average of multiple realizations of a sample size.
We use 20 realizations for sample sizes between 100,000 - 500,000, 10 realizations for samples between 600,000 - 900,000, and 5 realizations for sample sizes above 1,000,000 individuals.

}

\showmatmethods 

\acknow{V.S. and H.J would like to thank Sune Lehmann for useful discussions and feedback.}

\showacknow

\pnasbreak
\pnasbreak
\pnasbreak
\pnasbreak
\pnasbreak
\pnasbreak
\clearpage

\bibliography{biblio}

\begin{thebibliography}{10}

\bibitem{agrawal1993mining}
Agrawal R, Imieli\'{n}ski T, Swami A (1993) Mining association rules between
  sets of items in large databases.
\newblock {\em SIGMOD Rec.} 22(2):207--216.

\bibitem{bell2007lessons}
Bell RM, Koren Y (2007) Lessons from the netflix prize challenge.
\newblock {\em {ACM SIGKDD} Explorations Newsletter} 9(2):75--79.

\bibitem{chen2009large}
Chen Y, Pavlov D, Canny JF (2009) Large-scale behavioral targeting in {\em
  Proceedings of the 15th ACM SIGKDD international conference on Knowledge
  discovery and data mining}.
\newblock (ACM), pp. 209--218.

\bibitem{mislove2010you}
Mislove A, Viswanath B, Gummadi KP, Druschel P (2010) You are who you know:
  inferring user profiles in online social networks in {\em Proceedings of the
  third ACM international conference on Web search and data mining}.
\newblock (ACM), pp. 251--260.

\bibitem{dodds2010measuring}
Dodds PS, Danforth CM (2010) Measuring the happiness of large-scale written
  expression: Songs, blogs, and presidents.
\newblock {\em Journal of happiness studies} 11(4):441--456.

\bibitem{mislove2011understanding}
Mislove A, Lehmann S, Ahn YY, Onnela JP, Rosenquist JN (2011) Understanding the
  demographics of twitter users.
\newblock {\em ICWSM} 11:5th.

\bibitem{conte2012manifesto}
Conte R, et~al. (2012) Manifesto of computational social science.
\newblock {\em European Physical Journal-Special Topics} 214:p--325.

\bibitem{lazer2009computational}
Lazer D, et~al. (2009) Computational social science.
\newblock {\em Science} 323(5915):721--723.

\bibitem{blumberg2009locational}
Blumberg AJ, Eckersley P (2009) On locational privacy, and how to avoid losing
  it forever.
\newblock {\em Electronic Frontier Foundation} 10(11).

\bibitem{eckersley2010unique}
Eckersley P (2010) How unique is your web browser? in {\em International
  Symposium on Privacy Enhancing Technologies Symposium}.
\newblock (Springer), pp. 1--18.

\bibitem{de2013unique}
De~Montjoye YA, Hidalgo CA, Verleysen M, Blondel VD (2013) Unique in the crowd:
  The privacy bounds of human mobility.
\newblock {\em Scientific reports} 3:1376.

\bibitem{hannak2013measuring}
Hannak A, et~al. (2013) Measuring personalization of web search in {\em
  Proceedings of the 22nd international conference on World Wide Web}.
\newblock (ACM), pp. 527--538.

\bibitem{greenwood2014new}
Greenwood D, Stopczynski A, Sweatt B, Hardjono T, Pentland P (2014) The new
  deal on data: A framework for institutional controls.
\newblock {\em Privacy, big data, and the pubic good} pp. 192--210.

\bibitem{de2015unique}
De~Montjoye YA, Radaelli L, Singh VK, , et~al. (2015) Unique in the shopping
  mall: On the reidentifiability of credit card metadata.
\newblock {\em Science} 347(6221):536--539.

\bibitem{sapiezynski2015tracking}
Sapiezynski P, Stopczynski A, Gatej R, Lehmann S (2015) Tracking human mobility
  using wifi signals.
\newblock {\em PloS one} 10(7):e0130824.

\bibitem{mayer2016evaluating}
Mayer J, Mutchler P, Mitchell JC (2016) Evaluating the privacy properties of
  telephone metadata.
\newblock {\em Proceedings of the National Academy of Sciences} p. 201508081.

\bibitem{narayanan2008robust}
Narayanan A, Shmatikov V (2008) Robust de-anonymization of large sparse
  datasets in {\em Security and Privacy, 2008. SP 2008. IEEE Symposium on}.
\newblock (IEEE), pp. 111--125.

\bibitem{wiredhack}
Honan M (2012) {H}ow {A}pple and {A}mazon security flaws led to my epic hacking
  (\url{https://www.wired.com/2012/08/apple-amazon-mat-honan-hacking/}).

\bibitem{eagle2006reality}
Eagle N, Pentland AS (2006) Reality mining: sensing complex social systems.
\newblock {\em Personal and ubiquitous computing} 10(4):255--268.

\bibitem{stopczynski2014measuring}
Stopczynski A, et~al. (2014) Measuring large-scale social networks with high
  resolution.
\newblock {\em PloS one} 9(4):e95978.

\bibitem{itu2016}
Union IT (2016) International {Telecommunication} {Union}, {World}
  {Telecommunication}/{ICT} {Development} report and database.

\bibitem{stonetemple}
Enge E (2017) Mobile vs desktop usage: Mobile grows but desktop still a big
  player.
\newblock
  \newline\url{https://www.stonetemple.com/mobile-vs-desktop-usage-mobile-grows-but-desktop-still-
  a-big-player/}.

\bibitem{pagefair2017}
{PageFair} (2017) The state of the blocked web - 2017 global adblock report.
\newblock \newline\url{https://pagefair.com/blog/2017/adblockreport/}.

\bibitem{martin2018penalty}
Martin K (2018) The penalty for privacy violations: How privacy violations
  impact trust online.
\newblock {\em Journal of Business Research} 82:103--116.

\bibitem{posner1981economics}
Posner RA (1981) The economics of privacy.
\newblock {\em The American economic review} 71(2):405--409.

\bibitem{anthes2015data}
Anthes G (2015) Data brokers are watching you.
\newblock {\em Communications of the ACM} 58(1):28--30.

\bibitem{ramirez2014data}
Ramirez E, Brill J, Ohlhausen MK, Wright JD, McSweeny T (2014) Data brokers--a
  call for transparency and accountability.
\newblock {\em Federal Trade Commission, Tech. Rep}.
\newblock
  \url{www.ftc.gov/system/files/documents/reports/data-brokers-call-transparency-accountability-report-federal-trade-commission-may-2014/140527databrokerreport.pdf}.

\bibitem{sweeney2002k}
Sweeney L (2002) k-anonymity: A model for protecting privacy.
\newblock {\em International Journal of Uncertainty, Fuzziness and
  Knowledge-Based Systems} 10(05):557--570.

\bibitem{falaki2010diversity}
Falaki H, et~al. (2010) Diversity in smartphone usage in {\em Proceedings of
  the 8th international conference on Mobile systems, applications, and
  services}.
\newblock (ACM), pp. 179--194.

\bibitem{welke2016differentiating}
Welke P, Andone I, Blaszkiewicz K, Markowetz A (2016) Differentiating
  smartphone users by app usage.
\newblock {\em Proceedings of the 2016 ACM International Joint Conference on
  Pervasive and Ubiquitous Computing} pp. 519--523.

\bibitem{achara2015unicity}
Achara JP, Acs G, Castelluccia C (2015) On the unicity of smartphone
  applications in {\em Proceedings of the 14th ACM Workshop on Privacy in the
  Electronic Society}.
\newblock (ACM), pp. 27--36.

\bibitem{kosinski2013private}
Kosinski M, Stillwell D, Graepel T (2013) Private traits and attributes are
  predictable from digital records of human behavior.
\newblock {\em Proceedings of the National Academy of Sciences}
  110(15):5802--5805.

\bibitem{youyou2015computer}
Youyou W, Kosinski M, Stillwell D (2015) Computer-based personality judgments
  are more accurate than those made by humans.
\newblock {\em Proceedings of the National Academy of Sciences}
  112(4):1036--1040.

\bibitem{chittaranjan2013mining}
Chittaranjan G, Blom J, Gatica-Perez D (2013) Mining large-scale smartphone
  data for personality studies.
\newblock {\em Personal and Ubiquitous Computing} 17(3):433--450.

\bibitem{seneviratne2014predicting}
Seneviratne S, Seneviratne A, Mohapatra P, Mahanti A (2014) Predicting user
  traits from a snapshot of apps installed on a smartphone.
\newblock {\em ACM SIGMOBILE Mobile Computing and Communications Review}
  18(2):1--8.

\bibitem{malmi2016you}
Malmi E, Weber I (2016) You are what apps you use: Demographic prediction based
  on user's apps.
\newblock {\em Tenth International AAAI Conference on Web and Social Media}.

\bibitem{appbrain}
AppBrain (2017) {G}oogle {P}lay statistics
  (\url{https://www.appbrain.com/stats}).
\newblock Accessed: 2017-04-18.

\bibitem{dunbar1992neocortex}
Dunbar RI (1992) Neocortex size as a constraint on group size in primates.
\newblock {\em Journal of human evolution} 22(6):469--493.

\bibitem{alessandretti2016evidence}
Alessandretti L, Sapiezynski P, Lehmann S, Baronchelli A (2016) Evidence for a
  conserved quantity in human mobility.
\newblock {\em arXiv preprint arXiv:1609.03526}.

\bibitem{barbaro2006face}
Barbaro M, Zeller T, Hansell S (2006) A face is exposed for aol searcher no.
  4417749.
\newblock {\em New York Times} 9(2008):8For.

\bibitem{barth2012re}
Barth-Jones DC (2012) The're-identification'of {G}overnor {W}illiam {W}eld's
  medical information: a critical re-examination of health data identification
  risks and privacy protections, then and now.
\newblock {\em Available at SSRN: https://ssrn.com/abstract=2076397}.

\bibitem{sweeney2013identifying}
Sweeney L, Abu A, Winn J (2013) Identifying participants in the personal genome
  project by name.
\newblock {\em Available at SSRN: https://ssrn.com/abstract=2257732}.

\bibitem{tockar2014riding}
Tockar A (2014) Riding with the stars: Passenger privacy in the nyc taxicab
  dataset.
\newblock {\em Neustar Research, September} 15.

\bibitem{olmstead2016apps}
Olmstead K, Atkinson M (2016) Apps permissions in the {G}oogle {P}lay store.
\newblock {\em Pew Research Center}.

\bibitem{kossinets2006empirical}
Kossinets G, Watts DJ (2006) Empirical analysis of an evolving social network.
\newblock {\em Science} 311(5757):88--90.

\bibitem{sekara2016fundamental}
Sekara V, Stopczynski A, Lehmann S (2016) Fundamental structures of dynamic
  social networks.
\newblock {\em Proceedings of the national academy of sciences}
  113(36):9977--9982.

\bibitem{gdpr}
(2016) {Regulation (EU) 2016/679 of the European Parliament and of the Council
  of 27 April 2016 on the protection of natural persons with regard to the
  processing of personal data and on the free movement of such data, and
  repealing Directive 95/46/EC (General Data Protection Regulation)}.
\newblock {\em Official Journal of the European Union} L119:1--88.

\end{thebibliography}

\onecolumn
\section{Supplementary Information}
\renewcommand{\thefigure}{S\arabic{figure}}
\renewcommand{\thetable}{S\arabic{table}}
\setcounter{figure}{0}
\paragraph{The dataset}

We use a dataset that spans 12 months, from Feb. 1st 2016 to Feb. 1st 2017, and contains monthly app-fingerprints for 3,556,083 individuals.
Each fingerprint is a binary vector composed of the apps a person has used during a month. We do not consider apps that are installed but unused. 

We further disregard phone vendor specific apps such as: alarm clock, phone dialer, settings etc. and only focus on apps that are downloadable from Google Play.
This removes vendor bias, and makes re-identification harder. The users are selected from major markets in the Americas, Europe and Asia. Thus, the impact of regional variations on uniqueness due to local applications is smaller than if we had sampled users from anywhere in the world.

In total, the number of unique apps in the dataset is 1,129,110, and each individual in the dataset uses at least 3 apps per month.

The data was collected using a pre-loaded app recommender app on Xperia phones. Data collection is approved by the Sony Mobile Logging Board and written consent in electronic form has been obtained for all study participants according to the Sony Mobile Application Terms of Service and the Sony Mobile Privacy Policy.


\begin{figure}[!hptb]
  \centering
    \includegraphics[width=0.65\textwidth]{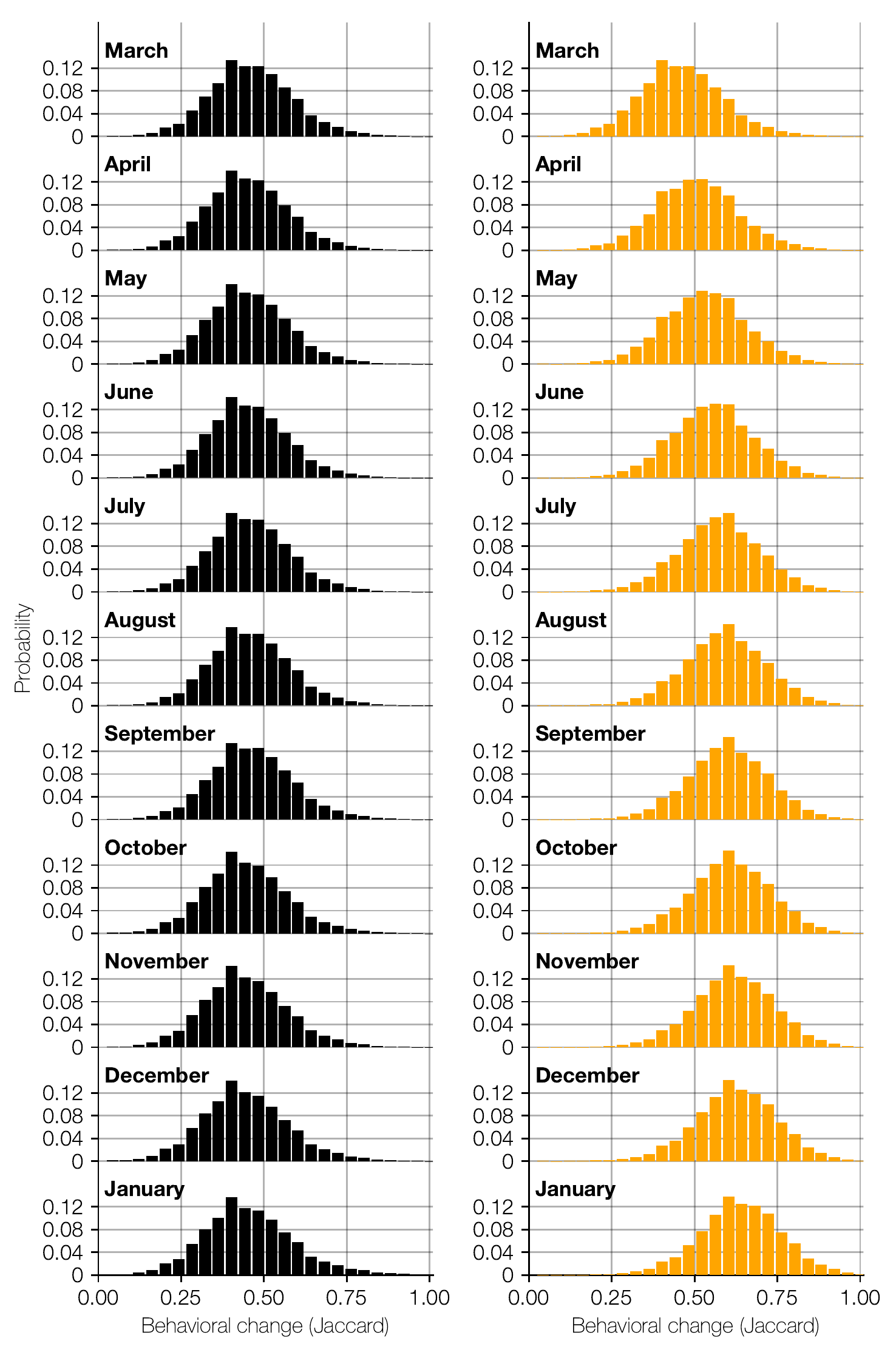}
    \label{fig:s1}
    \caption{Distributions show the change in app fingerprint over time. The change is measured as Jaccard distance between a users fingerprint in one month and the next. Left, change between consecutive months, e.g. February and March (denoted March), March and April (denoted April), etc. Right, Difference between fingerprint in February 2016 compared to other months, indicating a drift over time.}
\end{figure}


\begin{figure}[!hptb]
  \centering
    \includegraphics[width=0.8\textwidth]{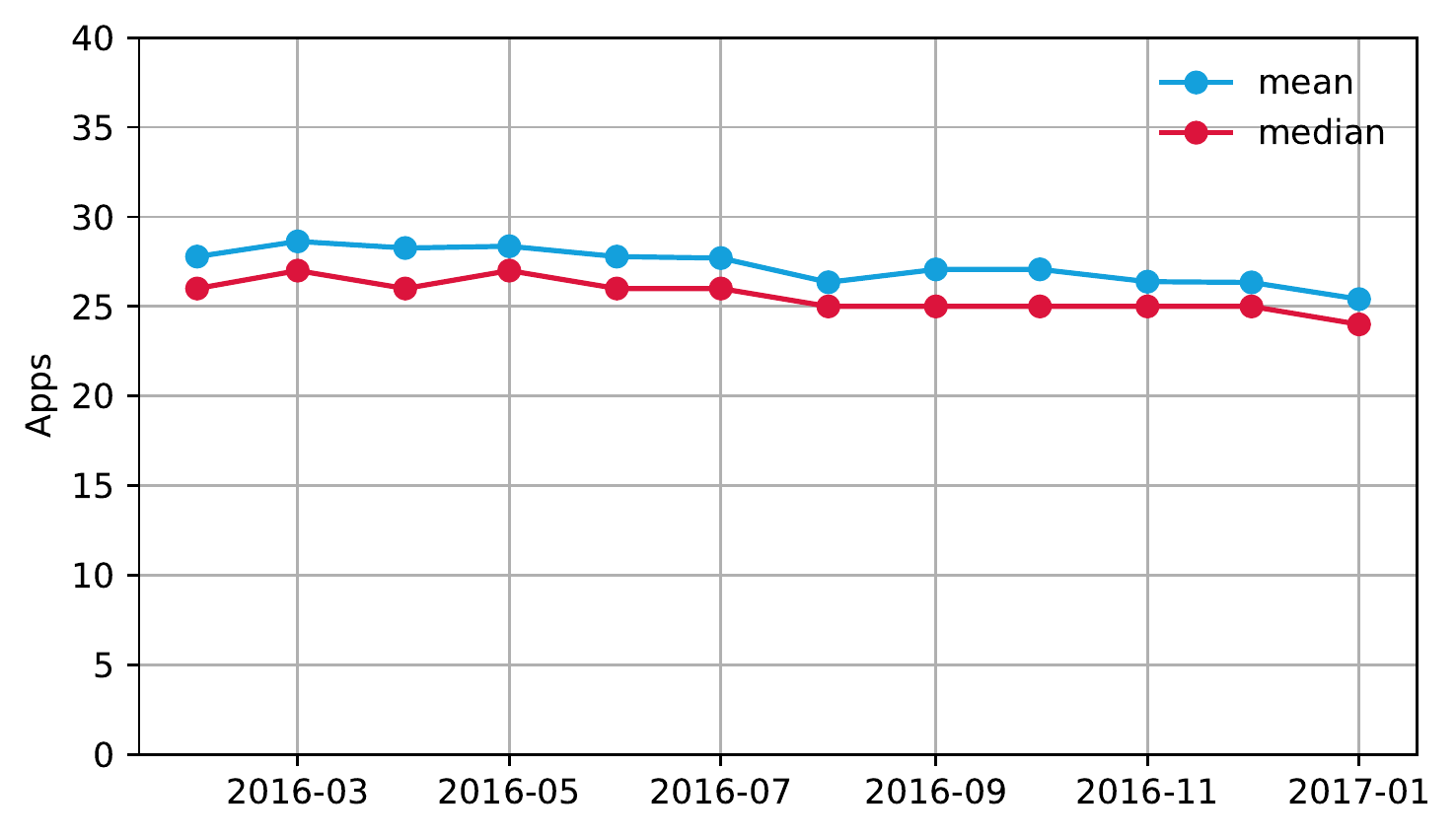}
    \label{fig:s2}
    \caption{Average number of apps per user per month. The median is also plotted for comparison.}
\end{figure}


\begin{figure}[!hptb]
  \centering
    \includegraphics[width=0.8\textwidth]{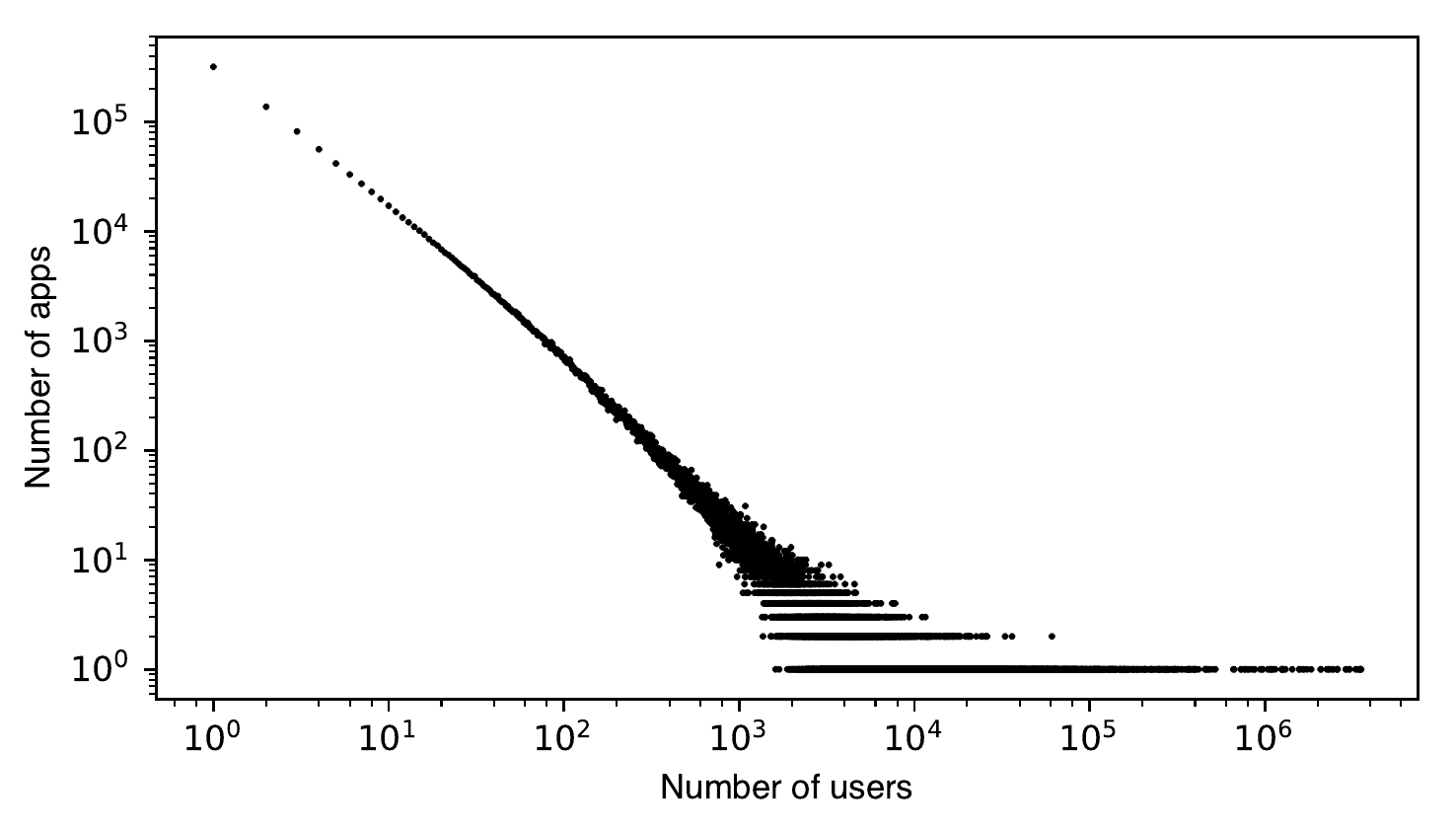}
    \label{fig:s3}
    \caption{Distribution of populatity of apps, i.e. the number of individuals using an app. Estimated across the entire dataset. Distribution clearly displays a long-tail. }
\end{figure}

\begin{figure}[!hptb]
  \centering
    \includegraphics[width=0.8\textwidth]{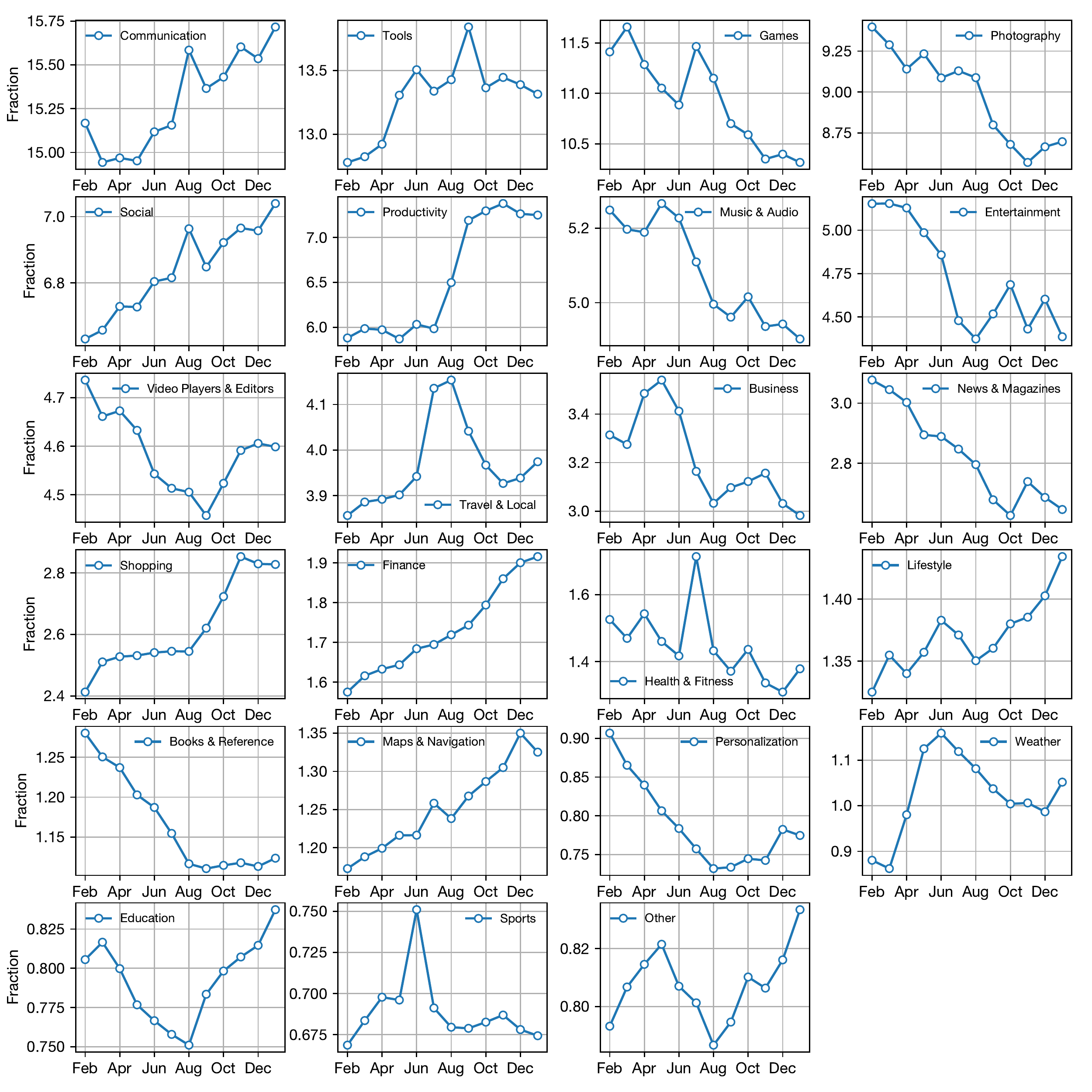}
    \label{fig:s4}
    \caption{Fraction of apps per category. Apps are divided into popular Google play categories and figure shows the fraction of app that belong to each category over time.}
\end{figure}


\begin{figure}[!hptb]
  \centering
    \includegraphics[width=0.8\textwidth]{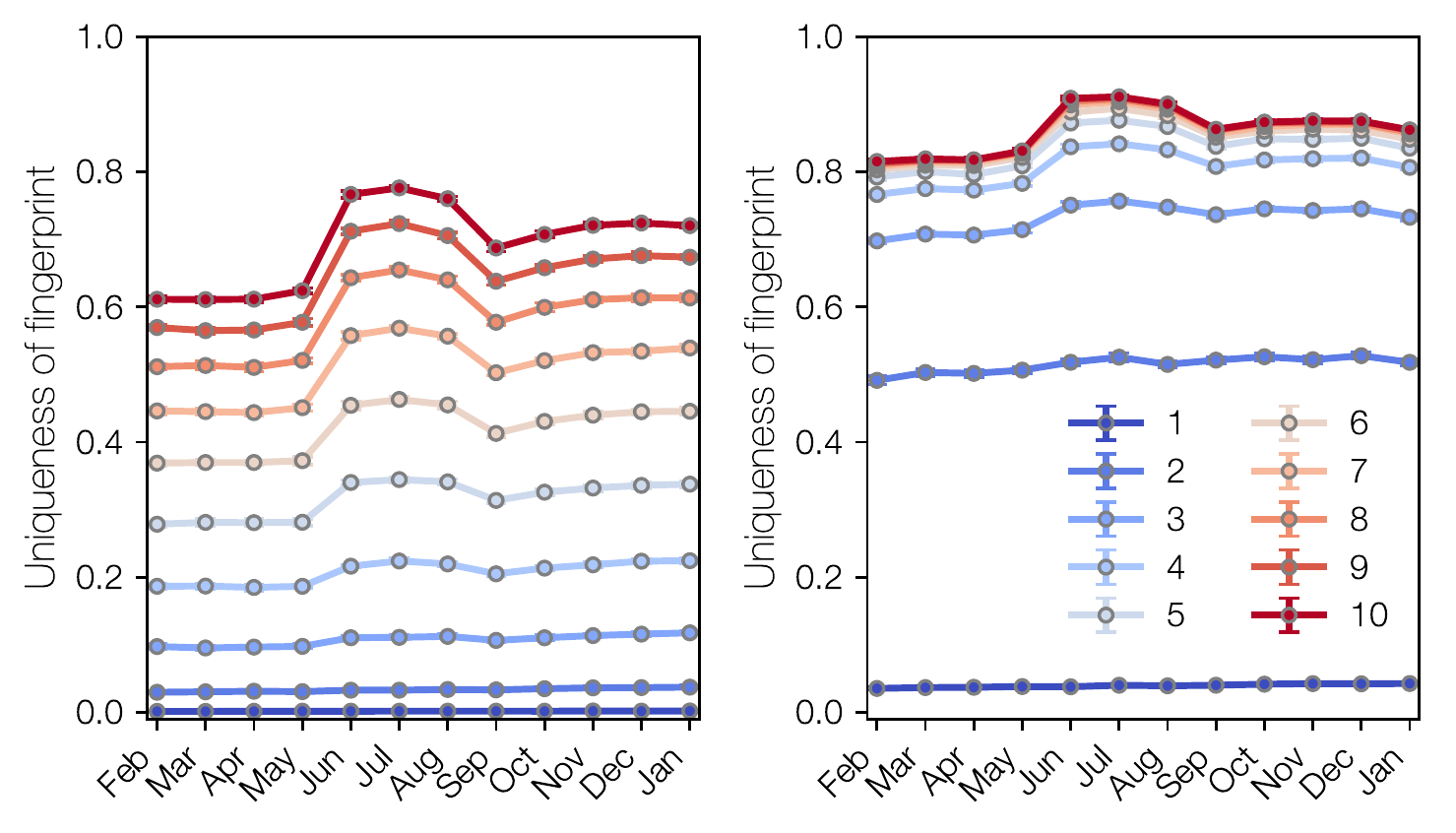}
    \label{fig:s5}
    \caption{Seasonable variations of uniqueness over time for the random scheme (left) and the popularity heuristic (right). Curves are rescaled according to $\widetilde{u}(t) = \frac{u(t)}{|A|_t/|A|_{t=0}}$, where $u(t)$ is the uniqueness at month $t$, and $|A|_t$ is the number of apps at time $t$. With $t=0$ denoting the first month of the dataset, February 2016. }
\end{figure}

\begin{table}[!hptb]
\centering
\begin{tabular}{lcccc}
\toprule
Function & Pseudo $R^2$& $a$ & $b$ & $\gamma$ \\
\midrule
$ax^{\gamma}+b$ & 0.939 & -0.031 & 0.989 & 0.504 \\ 
$a\exp(x^{\gamma})+b$ & 0.940 & -0.022 & 1.017 & 0.261 \\ 
$a\exp(\gamma x)+b$ & 0.931 & 0.066 & 0.914 & -0.388 \\ 
$ax + b$ & 0.908 & -0.014 & 0.975 & - \\ 
\bottomrule
\hline
\end{tabular}
\caption{Regression values.}
\label{table:s1}
\end{table}

\begin{figure}[!hptb]
  \centering
    \includegraphics[width=0.6\textwidth]{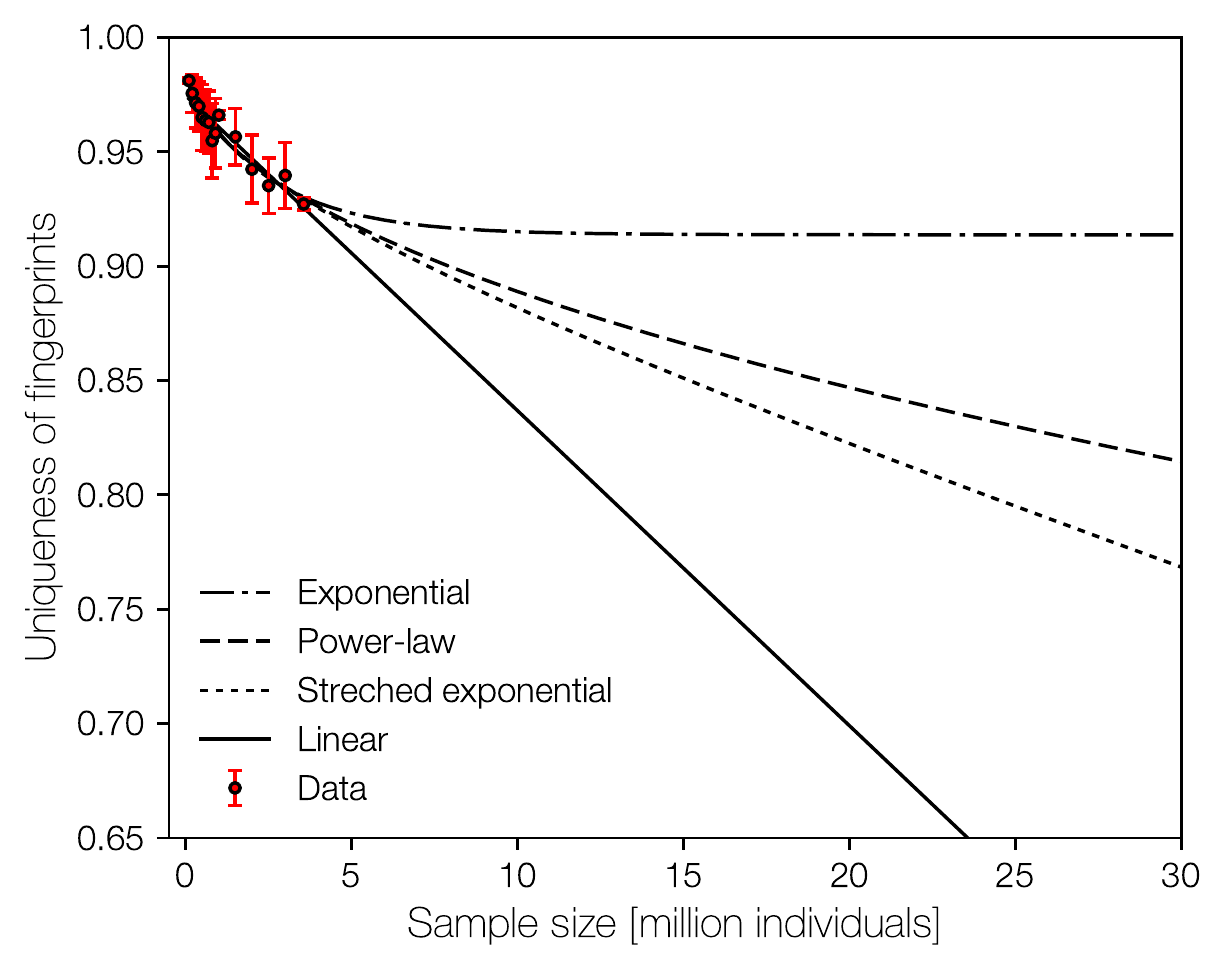}
    \label{fig:s10}
    \caption{Extrapolated uniqueness. Fit of different functional forms (see Table S1) to the uniqueness curve for $n_\text{apps} = 5$ when selecting apps using the popularity heuristic. Closes agreement with data is achieved by the stretched exponential and power law functional forms.}
\end{figure}

\end{document}